\documentclass[11pt]{article}
\usepackage{fleqn,cospar}

\usepackage{url}

\usepackage{graphicx}
\usepackage[figuresright]{rotating}

\newbox\grsign \setbox\grsign=\hbox{$>$} \newdimen\grdimen \grdimen=\ht\grsign
\newbox\simlessbox \newbox\simgreatbox \newbox\simpropbox
\setbox\simgreatbox=\hbox{\raise.5ex\hbox{$>$}\llap
     {\lower.5ex\hbox{$\sim$}}}\ht1=\grdimen\dp1=0pt
\setbox\simlessbox=\hbox{\raise.5ex\hbox{$<$}\llap
     {\lower.5ex\hbox{$\sim$}}}\ht2=\grdimen\dp2=0pt
\setbox\simpropbox=\hbox{\raise.5ex\hbox{$\propto$}\llap
     {\lower.5ex\hbox{$\sim$}}}\ht2=\grdimen\dp2=0pt
\def\simgt{\mathrel{\copy\simgreatbox}}

\def\tT{\tau_{\rm T}}
\def\Tbb{T_{\rm bb}}
\def\TC{T_{\rm C}}
\def\rtr{r_{\rm tr}}
\def\tK{t_{\rm K}}

\def\rcr{r_{\rm cr}}

\def\lcr{l_{\rm cr}}
\def\mus{\mu_s}

\def\LE{L_{\rm Edd}}
\def\dm{\dot{m}}
\def\dM{\dot{M}}

\newcommand{\bez}{\begin{eqnarray*}}
\newcommand{\eez}{\end{eqnarray*}}
\newcommand{\be}{\begin{equation}}
\newcommand{\ee}{\end{equation}}
\newcommand{\beq}{\begin{eqnarray}}
\newcommand{\eeq}{\end{eqnarray}}
\newcommand{\bc}{\begin{center}}
\newcommand{\ec}{\end{center}}

\title{ACCRETION DISK MODELS OF LUMINOUS BLACK HOLES}

\author{A.M. Beloborodov\address{Stockholm Observatory, 
    Saltsj\"obaden, SE-133 36, Sweden}}

\begin{document}

% typeset front matter
\maketitle

\begin{abstract}
Models of X-ray production in black-hole sources are reviewed and compared 
with recent observations. Possible diagnostics of the hot-disk and 
disk-corona models are discussed in the light of spectral and temporal data. 
A new model of a small-scale inviscid accretion disk is presented.
\end{abstract}

%#############################################################################

\section*{INTRODUCTION}

About 10\% of observed X-ray binaries are thought to harbor black holes 
(Tanaka \& Lewin 1995). In many cases the 
companion is a low-mass star and the black hole is fed by a gaseous 
stream pulled out from the companion through the Lagrangian $L_1$ point. 
The stream has high angular momentum and a large-scale rotating disk forms 
around the black hole. In 3 objects (Cyg~X-1, LMC~X-1, and LMC~X-3) the 
companion is a massive star with a strong wind. 
The wind-fed accretion flow is quasi-spherical with low angular momentum 
and can marginally form a small-scale disk.

A standard model was elaborated 3 decades ago for 
disk accretion onto galactic black holes (GBHs), 
see review by Pringle (1981). Owing to MHD instabilities the
rotating disk is turbulent and viscous
(see Balbus \& Hawley 1998 for a review). Viscosity dissipates the orbital 
energy and forces matter to spiral gradually towards the black hole. 
The bulk of energy is dissipated 
at a few Schwarzschild radii, $r_g=2GM/c^2=3\times 10^5(M/M_\odot)$~cm.
The maximum luminosity of the disk is of order of the Eddington limit, 
$\LE=2\pi r_gm_pc/\sigma_T=1.3\times 10^{38}(M/M_\odot)$~erg/s,
and the maximum blackbody temperature is 
$$
  kT_{\rm bb}^{\rm max}\sim \left(\frac{\LE}{\sigma r_g^2}\right)^{1/4}
  = 7\left(\frac{M}{M_\odot}\right)^{-1/4}{\rm ~keV}. 
$$
Similar accretion flows but of larger scales form around
super-massive black holes in AGN, $M\sim 10^8M_\odot$.
According to the $M^{-1/4}$ scaling AGN have $kT_{\rm bb}^{\rm max}\sim 70$~eV.

The puzzling property of GBHs and AGN is their hard X-ray emission
which sometimes dominates their spectra.
It indicates the presence of hot plasmas with temperature $kT\sim
100$~keV and scattering  optical depth $\tau_{\rm T}\sim 1$ near 
black holes (see reviews by Poutanen 1998, Zdziarski 1999).  
The plasma may be identified  with a hot two-temperature disk  
(e.g. Shapiro, Lightman \& Eardley 1976; Ichimaru 1977) 
or a corona atop a relatively cold disk
(e.g. Bisnovatyi-Kogan \& Blinnikov 1977;  
Galeev, Rosner, \& Vaiana 1979), see Beloborodov (1999b, hereafter B99b)
for a review. Recently, a new model of a small-scale inviscid disk was 
developed (Beloborodov \& Illarionov 2001, hereafter BI).

%#############################################################################

\section*{OBSERVATIONAL CHARACTERISTICS}

The X-ray sources classified as black holes have diverse spectral and
temporal properties. Black holes with low-massive companions  
are normally transient objects and generate outbursts with 
strong spectral evolution (see King, these proceedings). Persistent
high-mass sources may spend most of the time in the hard state dominated
by $\sim 100$~keV emission (Cyg~X-1) or in the soft state
dominated by $\sim 1$~keV quasi-blackbody peak (LMC~X-1, LMC~X-3). 

The measured X-ray spectra provide main information on the sources. 
In both GBHs and AGN
the spectrum is well represented as a sum of three components:
(1) the quasi-blackbody peak, (2) the intrinsic power-law 
emission from the X-ray source, and (3) the reflected/reprocessed X-rays
(including the Fe K$\alpha$ line and the Compton reflection bump, see 
Done, these proceedings). A couple of examples are shown in Fig.~4. 
The reflector is commonly identified with a (relatively cold, dense) disk. 
The origin of the hot X-ray source is less clear --
different models will be the main subject of this review.

The observed characteristics of the hot source are the photon index of the 
power law
$\Gamma$, the position of the spectrum break, and the amplitude of 
reflection $R$. The break is almost always at $50-200$~keV and 
$\Gamma$ varies from $\sim 2.5$ (soft spectra) to $\sim 1.5$ (hard spectra). 
The reflection amplitude measures the effective solid angle $\Omega$ subtended 
by the reflector as viewed from the X-ray source, $R=\Omega/2\pi$.
In most objects $R<1$ (see the $R-\Gamma$ diagram in Fig.~2).
Recently, a correlation was found between $R$ and $\Gamma$ in both 
GBHs and AGN (Zdziarski, Lubi\'nski, \& Smith 1999; Gilfanov, Churazov,
\& Revnivtsev 2000).

Additional information comes from the source variability. Variability is 
represented by the power density spectrum (PDS) of the fluctuating 
flux from the source (see Gilfanov, these proceedings). 
Typically, the power of variability peaks at time-scales $\sim 0.1-10$~s. 
Detailed studies of the PDS of Cyg~X-1 and GX~339--4 were recently performed 
with {\it RXTE} in different periods of the hard state. 
In the hardest periods, the variability shifted to longer time-scales 
(Gilfanov et al. 1999; Revnivtsev et al. 2000).

%###########################################################################

\section*{HOT DISK MODEL}

This model postulates the transition from the standard ``cold'' disk 
to a hot two-temperature ($T_p\gg T_e$) flow at some radius $\rtr$. 
The hot flow is a geometrically thick viscous rotating disk that accretes 
fast and has low density.
The mechanism of transition is not well established. One suggestion is that 
the cold disk may evaporate to a coronal flow atop it (e.g. 
R\'o\.za\'nska \& Czerny 2000).
The formation of the corona itself, however, depends on complicated MHD 
processes, and our understanding of these processes
is not sufficient to develop a unique transition model. For instance, 
it is unclear whether the corona stays always ``anchored'' by the magnetic 
field to the heavy cold disk or may form an independent accretion flow.
Another complicated issue is the heating mechanism 
and $e-p$ energy distribution in the viscous hot flow.
Given these uncertainties, different versions of the hot-disk model can be
developed. We will focus here on the phenomenological/observational aspect 
of the model.

Note that we discuss here {\it bright} sources (with luminosities
$L>10^{-2}\LE$) with substantial radiative efficiency of accretion.
Models with small efficiencies (advective flows) designed for low-luminosity 
sources are reviewed by Marek Abramowicz in these proceedings.
An intermediate regime with marginally important advection may apply
to the hard-state objects (Esin et al. 1998). It requires: (1) a large 
viscosity parameter $\alpha\sim 0.3$ and (2) an accretion rate 
just near the threshold for collapse of the hot flow into a cold disk.

\subsection*{Transition Radius and the X-Ray Spectrum}

The parameter $\rtr$ is especially important from the observational point of 
view. The energy released at $r>\rtr$, $L_d$, emerges as soft 
radiation according to the standard model (Shakura \& Sunyaev 1973). 
The energy released at $r<\rtr$, $L_X$, is dissipated in the hot flow that 
emits X-rays; a fraction of this energy can be stored in the flow and advected 
into the black hole (or away from it if an outflow forms).

The outer disk illuminates the central hot flow and cools it via 
Comptonization. For illustration, we will focus here on one version of this 
model where the outer disk is the main source of soft photons. Alternatively, 
the photons may be supplied by cold dense clouds embedded in the flow (e.g. 
Krolik 1998; Zdziarski et al. 1998). Besides, the hot flow itself can 
generate synchrotron photons (e.g. Esin et al. 1998; Wardzi\'nski \& Zdziarski
2000) and it becomes especially important if there are non-thermal particles 
whose emission is not self-absorbed. The observed $R-\Gamma$ correlation, 
however, suggests that the X-ray reprocessing by the disk dominates the
cooling radiation. 

The generated spectrum can be evaluated with a toy model 
of a homogeneous X-ray sphere of radius $R_X$
(Poutanen, Krolik \& Ryde 1997; Zdziarski et al. 1999; Gilfanov et al. 2000).
Let us fix $R_X=10r_g=20GM/c^2$ (the typical radius where the bulk of 
accretion energy is liberated) and consider $\rtr$ as a model parameter. 
If $\rtr<R_X$, the cold disk enters the X-ray source and $L_X$ decreases. 
If $\rtr>R_X$, there appears a ``gap'' between the outer cold disk and 
the X-ray emitter; then $L_X$ weakly depends on $\rtr$.

%%%%%%%%%%%%%%%%%%%%%%%%%%%%%%%%%%%%%%%%%%%%
\begin{figure*}[t]
\includegraphics[width=175mm]{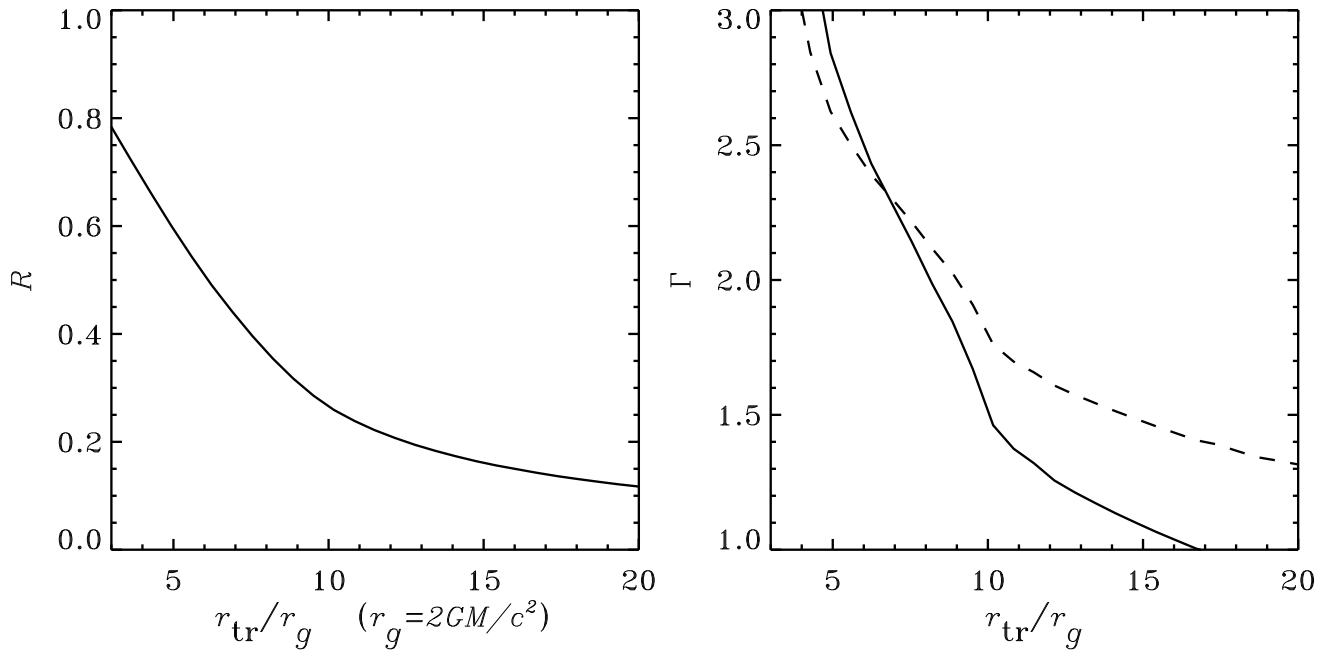}
\caption{The dependence of $R$ and $\Gamma$ on the transition radius. 
Solid and dashed curves correspond to GBHs ($\delta=1/6$) and AGN 
($\delta=1/10$) respectively (see eq.~1).}
\end{figure*}
%%%%%%%%%%%%%%%%%%%%%%%%%%%%%%%%%%%%%%%%%%%%

The luminosity $L_X$ is partly reprocessed/reflected from the cold disk.
The albedo $\sim 0.1$ for a neutral reflector; then 90\% of the X-rays 
impinging the disk are reprocessed into blackbody soft radiation. 
The observed blackbody component has luminosity $L_{\rm bb}=L_d+L_{\rm repr}$
where $L_{\rm repr}$ is the reprocessed luminosity.
Using the standard Monte-Carlo technique we compute $L_{\rm bb}$ for
given accretion rate and $\rtr$, and then find the soft luminosity entering the 
X-ray sphere, $L_s$.  In a steady state, 
the Compton amplification factor of the sphere is $A=(L_X+L_s)/L_s$.
The spectral index of the Comptonized radiation is related to $A$ (B99b)
\begin{equation}
   \Gamma=\frac{7}{3}A^{-\delta},
\end{equation}
where $\delta=1/10$ for AGN and $\delta=1/6$ for GBHs.
The resulting $\Gamma(\rtr)$ is shown in Fig.~1. 
Note that relevant $\Gamma$ are obtained in a narrow range $5<\rtr/r_g<15$.
Large $\rtr$ are excluded: then a very small fraction
of $L_{\rm bb}$ enters the inner X-ray source, leading 
to photon starvation and too hard spectrum of the source
(this conclusion relies, however, on the assumption that the outer disk is 
the dominant source of soft radiation). 

From the simulation we also determine 
the reflection amplitude $R(\rtr)$. We take into account the attenuation
of the observed reflection component by scattering in the hot sphere 
(hereafter $R$-attenuation). In the calculations the $R$-attenuation 
is estimated assuming Thomson optical depth of the sphere $\tT=1$.

Both $R$ and $\Gamma$ decrease with $\rtr$. The resulting track on the
$R-\Gamma$ diagram is shown in Fig.~2 (solid curve). 
The model predicts low reflection amplitudes throughout the whole range of 
spectral indexes. Note that the reflector was assumed to be neutral. 
If the disk is strongly ionized then reflection is further suppressed.

%%%%%%%%%%%%%%%%%%%%%%%%%%%%%%%%%%%%%%%%%%%%
\begin{figure*}
\includegraphics[width=172mm]{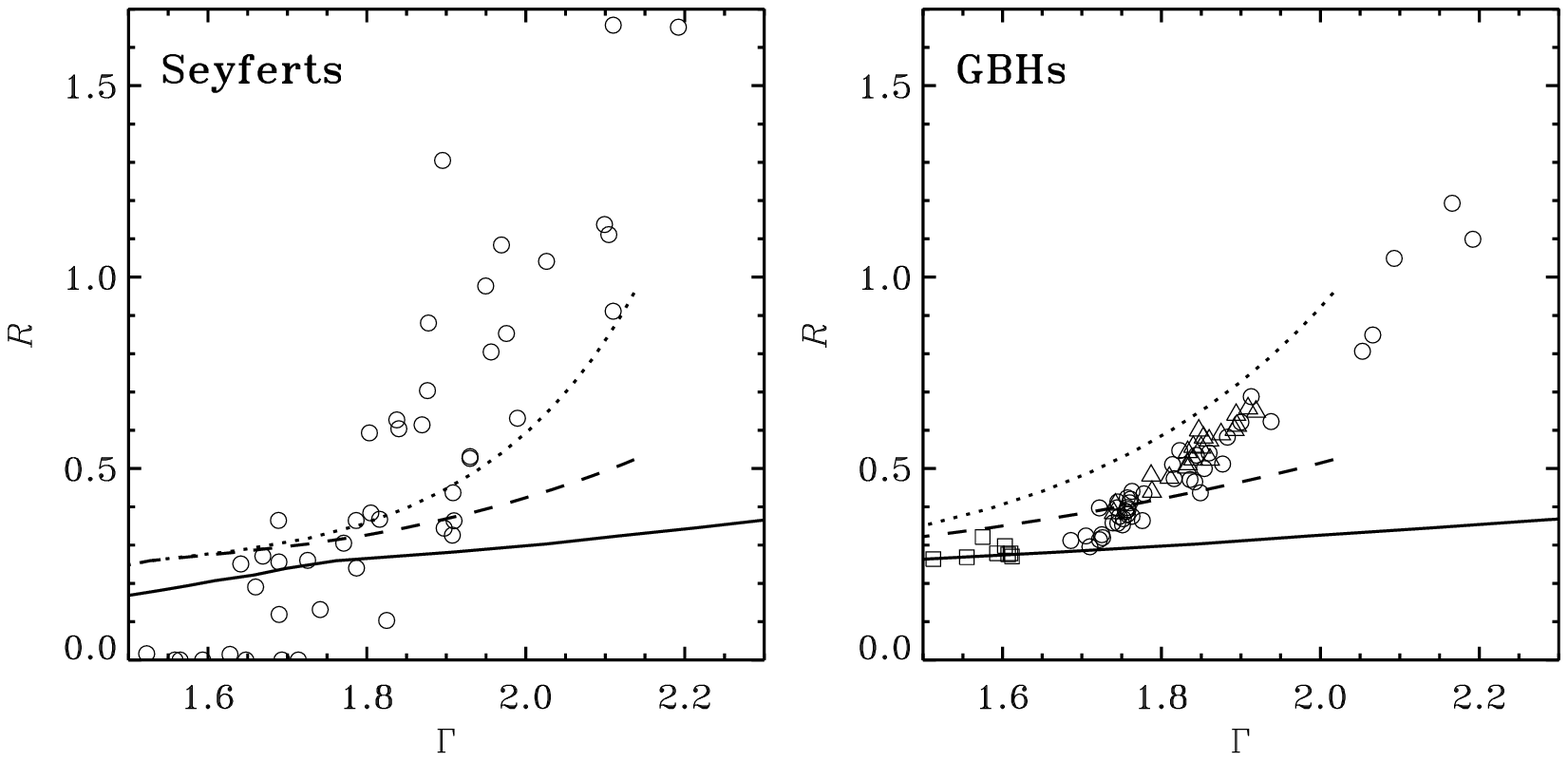}
\begin{minipage}{175mm}
{\sf Fig.~2. $R-\Gamma$ diagram (data from Zdziarski et al. 1999,
Gilfanov et al. 2000). Symbols show the best-fit values (large
errors are not shown here, see Zdziarski et al. 1999). 
Solid curve shows the model computations for the
hot disk. The main cooler of the X-ray source turns out to be the 
intrinsic radiation of the cold disk rather than reprocessed radiation.
To emphasize the importance of intrinsic radiation,
we also show the model where only 
reprocessed radiation cools the X-ray sphere (dashed curve).
Dotted curve shows the model where both the intrinsic luminosity of the 
cold disk and the $R$-attenuation effect are neglected.}
\end{minipage}
\end{figure*}
%%%%%%%%%%%%%%%%%%%%%%%%%%%%%%%%%%%%%%%%%%%%

\subsection*{Variability}

The energy release can be unsteady in the inner hot flow, leading to
variable emission. The shortest time-scale in the disk is the Keplerian 
period, $\tK\sim 3\times 10^{-2}(r/10r_g)^{3/2}(M/10M_\odot)$~s.
It is also the typical time-scale for
MHD instabilities. On longer time-scales the disk emission is averaged over 
many rotations and the observed variations may be associated with unsteady
radial dynamics of the flow. The radial (accretion) time-scale is 
$t_a\sim(\tK/2\pi\alpha)\sim 0.1(\alpha/0.2)^{-1}$s 
(assuming a thick disk with height $H\sim r$), 
so the hot disk accretes fast compared to typical observed variability.
 
Modulations may also come from the outer disk. E.g. $\rtr$ may 
decrease on the accretion time-scale of the outer disk, leading to an increase 
in $R$ and $\Gamma$. Accretion is relatively
slow in the standard disk and 
the rate of variations at $r\sim 10r_g$ is
consistent with the observed time-scales of maximum variability (see below).

%###########################################################################

\section*{DISK-CORONA MODEL}

In this model, the dense thin disk extends down to the marginally stable 
orbit and the X-ray emission is attributed to its coronal activity.
Hot coronae are believed to form as a result of
magnetorotational  instabilities  in  the  disk  and  the  buoyancy  of the
generated  magnetic field  (Tout \& Pringle 1992; Miller \& Stone 2000).  
The corona is heated in flare-like events of magnetic
dissipation producing the variable X-ray emission. There is no unique
model for the coronal activity; like the hot-disk model, 
the parameters of the putative corona are inferred from observations.

The two main parameters are: (1) the fraction of viscous energy that is 
released in the corona, $f$, (the remaining fraction, $1-f$, is dissipated 
inside the disk). Observations require $f$ up to $\sim 50$\% and similar 
values are suggested by the MHD models. (2) The feedback factor $D=L_s/L$
that is the fraction  of  the  X-ray luminosity that is reprocessed {\it and}
reenters  the  X-ray source
(Haardt \& Maraschi 1993; Stern et al. 1995; Poutanen \& Svensson 1996).
The geometry of the X-ray corona is unknown:
it may be e.g. a large cloud covering the whole inner region of the disk or 
a number of small-scale blobs (flares) with short life-times.
From the spectral point of view, the only important parameter of the 
geometry is the effective feedback $D$ that regulates the temperature of the 
corona and determines its equilibrium Compton amplification factor,
\begin{equation}
  A=\frac{L}{L_s}=\frac{1}{D}.
\end{equation}
This equation states the energy balance of the corona 
(here, for simplicity, the intrinsic flux from the disk has been neglected 
compared to the reprocessed flux; this is the case for intense concentrated 
flares).

\subsection*{Static Corona with Neutral Reflector}

Most of the previous computations of the disk-corona spectra assumed a
static corona located atop a neutral reflector (see reviews by
Svensson 1996, Poutanen 1998). The model was successfully applied to a number 
of Seyfert~1s. However, it was found to disagree with observations of  
black-hole sources in the hard state, e.g. Cyg X-1 (Gierli\'nski et al. 1997). 
The model
never predicts small $R$ simultaneously with hard spectra, in particular
$R\sim 0.3$ and  $\Gamma\sim  1.6$ observed in Cyg~X-1 cannot be reproduced.
Furthermore,  changes in the coronal geometry produce an anticorrelation 
between $\Gamma$ and $R$ (Malzac, Beloborodov, \& Poutanen 2001, hereafter
MBP). This anticorrelation is opposite to what is observed.

\subsection{Static Corona with Ionized Reflector}

The X-rays from concentrated flares can strongly ionize the upper layers of 
the disk. In the limit of complete ionization, the disk would resemble a 
perfect mirror at energies $h\nu<20$~keV: a power-law incident spectrum is 
reflected into a power-law without any reflection features and an observer 
will derive $R=0$. Ionization substantially reduces the observed amplitude 
of reflection when the ionization parameter $\xi\simgt 10^3$ (e.g. \.Zycki 
et al. 1994; Ross, Fabian, \& Young 1999; Nayakshin, Kazanas, \& Kallman 
2000).

Approximately, the ionized ``skin'' of the disk can be represented
as a completely ionized layer at a Compton temperature $k\TC\sim 10$~keV atop 
a neutral reflector. 
With increasing ionization parameter, the optical depth of the skin
$\tau_s$ increases and the amplitude of refection is suppressed. Also the 
reprocessed radiation is suppressed, leading to a harder spectrum of the 
corona. The hardness of the spectrum leads to even larger $\tau_s$ 
(Done \& Nayakshin 2001).

For illustration consider a slab corona atop optically thick neutral material 
and a skin $\tau_s$ between them. The X-ray flux illuminating the 
skin, $F_i$, is mostly reflected after a few scatterings, and a 
fraction $F_1\sim(1+\tau_s)^{-1} F_i$ reaches the neutral material.
The neutral material reflects a fraction $a\sim 0.1$ of $F_1$ and the rest
is reprocessed into blackbody radiation. 
The blackbody component then diffuses out of the skin with a ``mirror'' inner 
boundary condition and emerges with flux $\approx F_1$. The reflection 
component diffuses with an absorbing inner boundary and it gets additionally 
suppressed by a factor $\sim (1+\tau_s)^{-1}$. Thus the skin suppresses 
$R$ as $\sim (1+\tau_s)^{-2}$ and $L_s$ as $\sim(1+\tau_s)^{-1}$.

These estimates however do not work for a patchy corona. In that case
one needs a 2D model: the disk is strongly ionized beneath the flare
and the ionization parameter (and $\tau_s$) will decrease outside
the flare. The reflection features (and possibly the bulk of soft radiation)
then comes from the less ionized region around the flare, not from beneath it.

\subsection*{Dynamic Corona}

The static model is not self-consistent for $e^\pm$-dominated flares because 
the created pairs immediately (on Compton time-scale $\ll$ light-crossing time)
acquire the equilibrium bulk velocity $\beta=v/c\sim 0.5$ in the anisotropic
radiation field (Beloborodov 1999a, hereafter B99a). Besides, an anisotropic 
energy input of the flare must eject the hot plasma like the ejection in solar 
flares. The plasma can be ejected away or towards the disk with the 
preferential direction away from the  disk. 
One can show that bulk acceleration is also efficient for normal $e-p$ plasma:
$\beta>0.1$ in flares with compactness parameters $l>10^2$ (B99b). 
Note also that the proton component may be hot, with about virial temperature,
and then the plasma is likely to outflow with a virial velocity $\beta>0.1$.

Mildly relativistic bulk motion in the X-ray source causes X-ray
aberration and strongly affects the observed $R$ and $\Gamma$.  
B99a estimated the dependence of $R$ and $\Gamma$ on $\beta$ and found that 
$R\sim  0.3$ and $\Gamma\sim  1.6$ in Cyg~X-1 can both be explained
with a neutral reflector assuming that the emitting plasma outflows with
$\beta\approx 0.3$.
Recently, MBP performed exact Monte-Carlo computations  of the X-ray spectra
produced  by dynamic coronae and confirmed the simple analytical model
of B99a. In particular, the spectrum of Cyg~X-1 was well modeled with 
$\beta=0.3$ (see Fig.~4).

Fig.~3 illustrates  the  effects  of bulk  motion on the emitted spectra. 
In the case of $\beta=0.3$  (plasma moves upwards) 
the X-rays are beamed away from the disk; as a result the apparent X-ray
luminosity is enhanced while the reprocessed and reflected
luminosities  are  reduced.  The low  feedback  leads  to a hard spectrum. 
In the case of $\beta=-0.2$ (plasma moves downwards) the
X-rays are beamed towards the disk and the  reprocessed and reflected 
components are enhanced. The high feedback leads to a soft spectrum.

%%%%%%%%%%%%%%%%%%%%%%%%%%%%%%%%%%%%%%%%%
\begin{figure*}
\begin{center}
\includegraphics[width=175mm]{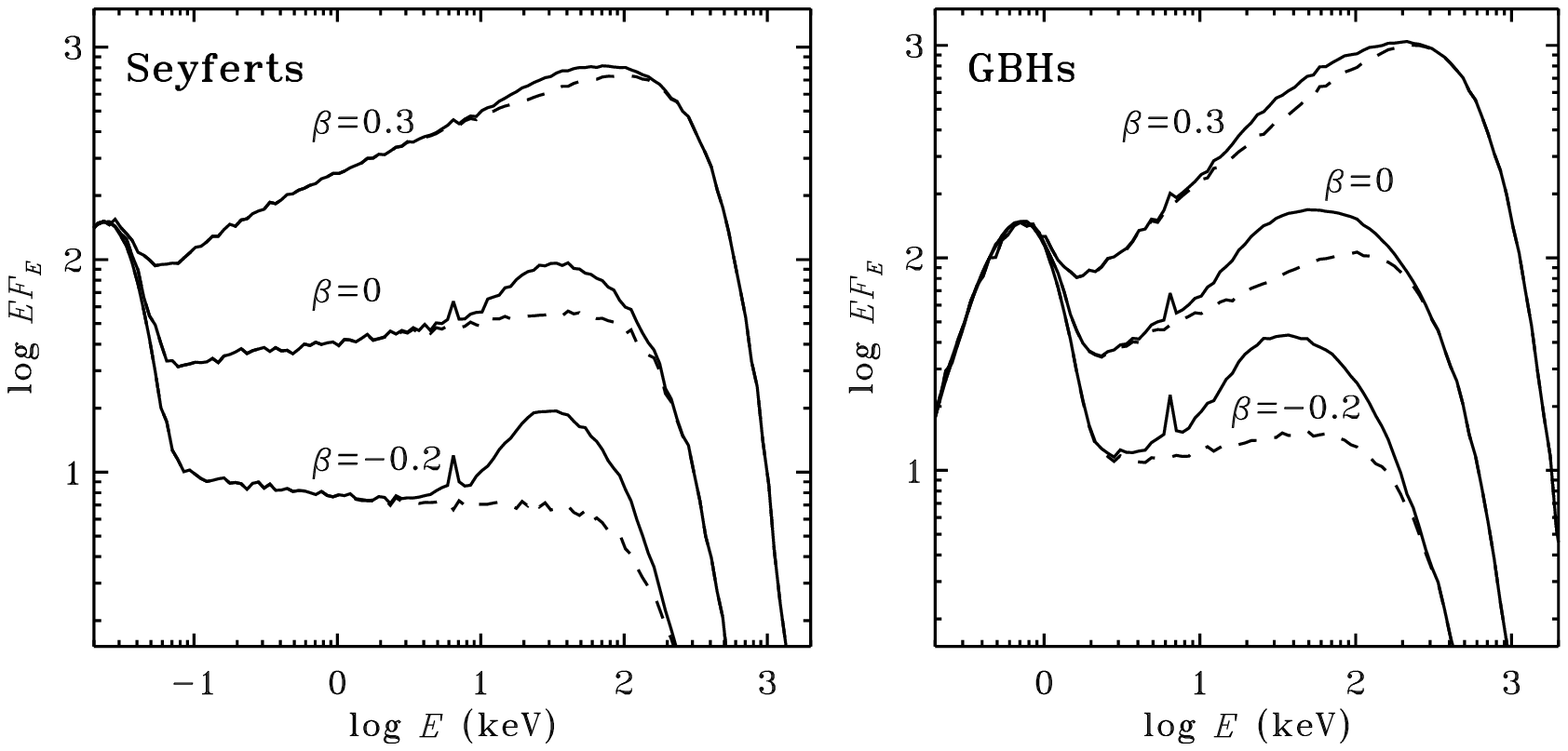}
\end{center}
\begin{minipage}{175mm}
{\sf Fig.~3. Effect  of bulk  motion  on the  emitted  spectra.
Here $h/r=2$, $\tT=3$, and nearly face-on inclination is assumed,
$0.9<\cos i<1$. {\em (Left)\/} AGN ($k\Tbb=5$~eV). {\em (Right)\/} GBHs 
($k\Tbb=150$~eV). (From MBP.) }
\end{minipage}
\end{figure*}
%%%%%%%%%%%%%%%%%%%%%%%%%%%%%%%%%%%%%%%%%

MBP computed the X-ray spectra from hot cylinders with height to radius
ratio $h/r$ taken as a parameter.
The observed reflection amplitude can be approximated by formula
\beq
R(\mu)=\frac{(1-\beta\mu)^3}{(1+\beta\mus)^2}
  \left\{\mus\left(1+\frac{\beta\mus}{2}\right) 
 +\frac{(1-\mus)\left[1+\beta(1+\mus)/2\right]}{(1+\beta)^2}\;
  e^{-\tT(1-\mus)}
  \right\},
\eeq
where  $\mu=\cos i$, $i$ is the  inclination  angle of the disk, and the 
parameter $\mus\approx(h/2r)/\sqrt{1+(h/2r)^2}$ describes the flare geometry.
The reflected luminosity is here represented as a sum of two parts: 
(1) reflected outside the cylinder base, which does not experience any 
attenuation, and (2) reflected from the base, which is attenuated
depending on $\tT$. MBP show that formula~(3) is in excellent agreement
with exact simulations.

%%%%%%%%%%%%%%%%%%%%%%%%%%%%%%%%%%%%%%%%%
\begin{figure*}
\begin{center}
\includegraphics[width=175mm]{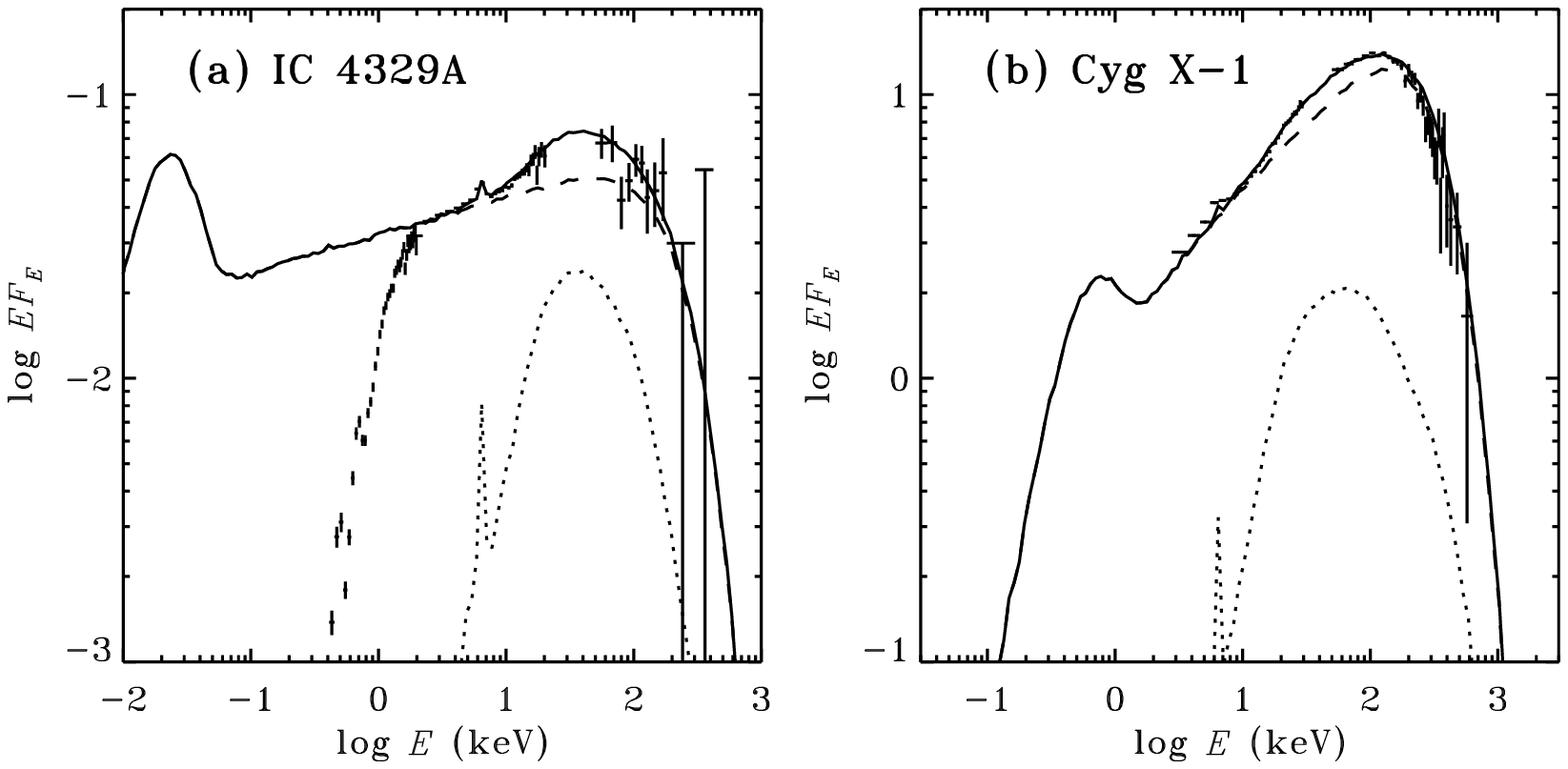}
\end{center}
\begin{minipage}{175mm}
{\sf Fig.~4. {\em (Left)\/} Spectrum of the Seyfert 1 galaxy IC4329A 
(Madejski et al. 1995). Solid curve is the model spectrum for $\tT=3$, $h/r=2$,
$\beta=0.1$ at inclination $i=40^{\rm o}$. {\em (Right)\/} Spectrum of Cyg~X-1 
observed in September 1991 (Gierli\'nski et al. 1997).
Solid curve is the model spectrum for $\tT=3$, $h/r=1.25$,
$\beta=0.3$ at $i=50^{\rm o}$.
In both panels, dotted curves display the reflected components,
dashed curves show the intrinsic Comptonized spectra. (From MBP.)}
\end{minipage}
\end{figure*}
%%%%%%%%%%%%%%%%%%%%%%%%%%%%%%%%%%%%%%%%%

%%%%%%%%%%%%%%%%%%%%%%%%%%%%%%%%%%%%%%%%%
\begin{figure*}
\begin{center}
\includegraphics[width=175mm]{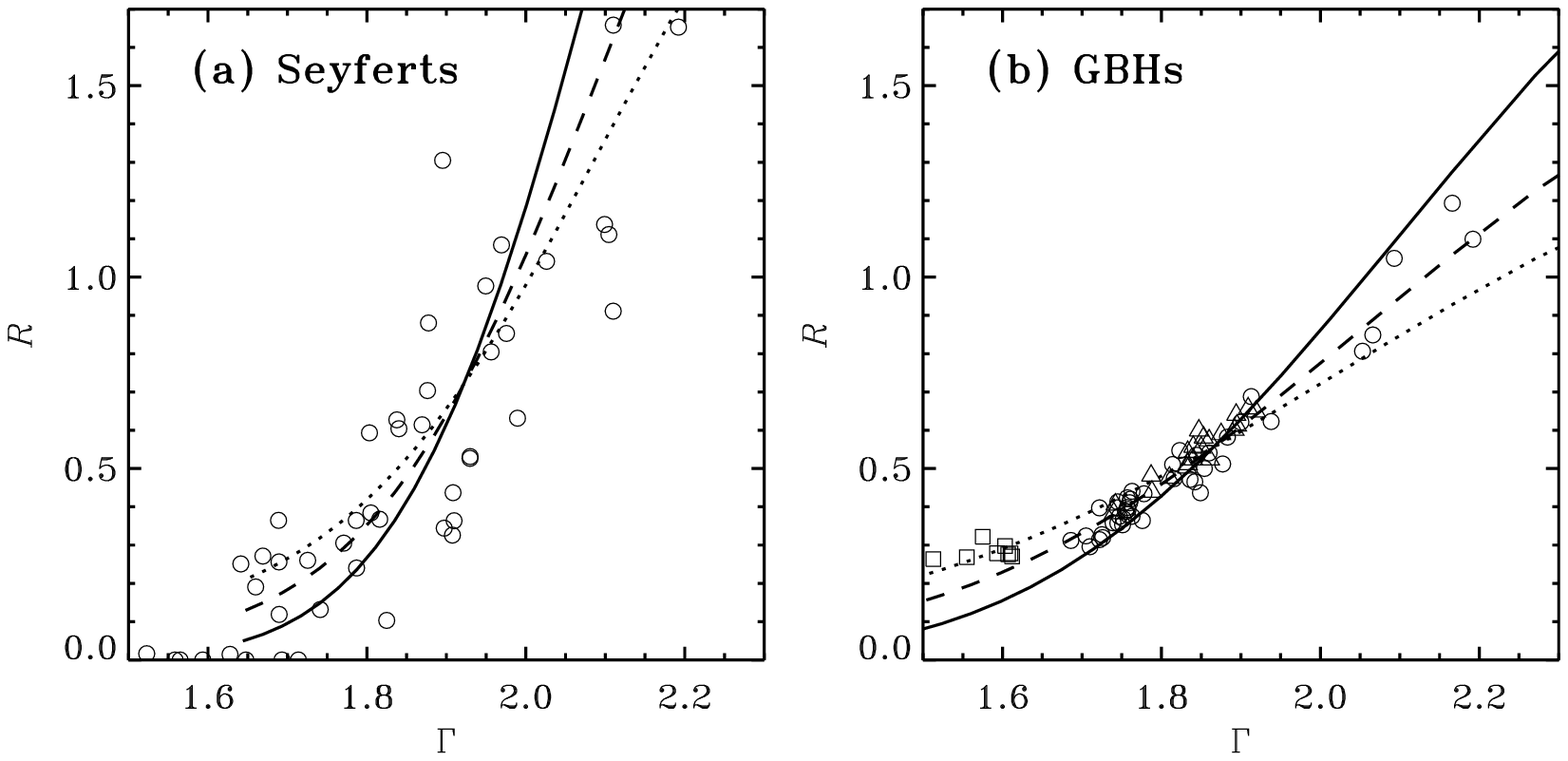}
\end{center}
\begin{minipage}{175mm}
{\sf Fig.~5. {\em (Left)\/} $R$--$\Gamma$ diagram for
Seyfert galaxies; data from Zdziarski et al. (1999).
The curves show the dynamic model with
$\mu_0=0.6$, $\tT=3$, $a=0.15$ at three different inclinations.
{\em (Right)\/} $R-\Gamma$ diagram for GBHs: Cyg~X-1 (circles), GX~339--4 
(triangles) and GS 1354-644 (squares); data from Gilfanov et al. (2000).
The model curves have $\mu_0=0.45$, $\tT=3$, $a=0.15$.
In both panels, solid, dashed, and dotted curves correspond to
$i=30^{\rm o}, 60^{\rm o}$, and $70^{\rm o}$, respectively. (From MBP.)}
\end{minipage}
\end{figure*}
%%%%%%%%%%%%%%%%%%%%%%%%%%%%%%%%%%%%%%%%%

The spectral index of the source, $\Gamma$, can be evaluated using 
relation (1) combined with the formula for the amplification factor (B99a),
\be \label{eq:A}
  A=\frac{2\gamma^2(1+\beta)^2(1+\beta\mus)^2}
         {(1-a)(1-\mus)[1-\beta^2(1+\mus)^2/4]},
\ee
where $a$ is the energy-integrated albedo of the disk.
Equations (1,3,4) give $R(\beta)$ and $\Gamma(\beta)$ predicted by the 
dynamic corona model for given $\mus$, $\tT$, $a$. With increasing $\beta$, 
$R$ and $\Gamma$ decrease. The resulting track in the $R-\Gamma$ diagram 
is consistent with observations (see Fig.~5).
 
The important specific prediction of the dynamic corona model is that the 
scattered soft radiation acquires strong polarization parallel to the disk 
normal (Beloborodov 1998b), in agreement with optical polarimetric 
observations of Seyferts (Koratkar \& Blaes 1999). Further diagnostics
is possible in case the disk emission has a Lyman edge in absorption.
Then the observed flux bluewards of the edge should be dominated by
the polarized upscattered radiation from the corona. This is
consistent with the recently discovered steep rise in polarization bluewards 
of the Lyman limit (see Beloborodov \& Poutanen 1999 and references therein).

\subsection*{Variability}

Like the hot-disk model, there is no physical model for variability
of coronae (but see Poutanen \& Fabian 1999 for a phenomenological 
model reproducing the observed variability).
The life-time of an individual flare is short
(probably comparable to Keplerian time, $\tK\sim 10$~ms at $r\sim 10r_g$). 
The observed
variability on longer time-scales can be attributed to changes in 
the accretion disk dynamics, in particular to variations in the magnetic 
structure of the disk. A possible time-scale for such variations is
the accretion time,
$$
 t_a\sim \frac{r_g}{\alpha c}\left(\frac{r}{r_g}\right)^{3/2}
         \left(\frac{r}{H}\right)^2
    \sim \frac{r_g}{\dm^2\alpha c}\left(\frac{r}{r_g}\right)^{7/2}
    \sim \frac{3}{\dm^2}\left(\frac{M}{10M_\odot}\right)
         \left(\frac{0.1}{\alpha}\right)
         \left(\frac{r}{10r_g}\right)^{7/2}{\rm s},
     \quad \dm\equiv\frac{\dot{M}}{\LE c^2}=\frac{17.5L}{\LE}.
$$
Here we assumed a non-rotating black hole and a radiation-dominated disk 
($L>10^{-2}\LE$). At $r\sim 10r_g$
this time-scale is consistent with the peak in observed PDSs, and
smaller radii may be responsible for variability at higher frequencies.
The flare ejection velocity probably increases (and $R$ and $\Gamma$ decrease)
at smaller $r$. This picture may explain the results of Fourier-resolved 
spectroscopy (Gilfanov et al. 2000).

Transitions between different spectral states may be triggered by
variations of e.g. $\dM$.
Observations indicate that when the disk goes to the hard state 
the intensity of coronal dissipation $f$ 
increases and an outflow forms. The latter is confirmed by radio observations:
radio-jet highly correlates with the hard state (Fender 2000).
At the same time, typical time-scales of flux fluctuations increase. 
It can be naturally explained by deceleration of accretion: the larger $f$
the thinner the cold disk (Svensson \& Zdziarski 1994) and $t_a$ grows 
$\propto (r/H)^2$. If the outflow is not advective, so that the accretion 
efficiency stays high in the hard state,
then the observed flux should be enhanced by beaming at $\dM=const$. However,
the actual behavior of $\dM$ is not known and it possibly decreases in
the soft-to-hard transition. This can be tested in high-inclination objects
where the observed flux should decrease with decreasing $\dM$.

%###########################################################################

\section*{SMALL-SCALE INVISCID DISK}

\subsection*{Quasi-Spherical Accretion in Bright Sources}

In many sources the accretion flow can be quasi-spherical, with low
angular momentum $l$, so that the disk formation is marginal. 
This is known to be the case in wind-fed X-ray binaries
(Illarionov \& Sunyaev 1975a,b; Shapiro \& Lightman 1976).
The net specific angular momentum of the trapped wind matter is
$\bar{l}_z\sim (1/4)\Omega R_a^2$ 
where $R_a=2GM/w^2\sim 10^{11}$~cm is the accretion
radius, $w\approx 10^8$~cm~s$^{-1}$ is the wind velocity, and $\Omega$ 
is the binary angular velocity. From the observed $\Omega$ one finds
$\bar{l}_z\sim r_gc$ for all the three known X-ray binaries classified as 
black holes with massive companions (Cyg~X-1, LMC~X-1, and LMC~X-3, see 
Tanaka \& Lewin 1995).
The accreting wind matter then inflows quasi-spherically at radii 
$r_g\ll r<R_a$ with a marginal formation of a small-scale disk at $r\sim r_g$.
Similar accretion flows may form around massive black holes in AGN. 
Among possible gas sources in AGN are star-star collisions and tidal 
disruption of stars by the black hole (Hills 1975); the accreting gas then 
likely has modest angular momentum.

Quasi-spherical flows in luminous sources must be Compton cooled and fall 
freely towards the black hole along ballistic trajectories 
(Zel'dovich \& Shakura 1969). The threshold condition for disk formation 
reads $l>l_*=0.75r_gc$: then rotation deflects the trajectories from the 
radial direction and a ring-like caustic appears in the defocused inflow 
outside the black hole horizon. Here matter liberates orbital energy in 
inelastic collision and then proceeds via a thin disk into the black hole. 
If the inflow has $\bar{l}_z$ comparable to $l_{\rm ms}=\sqrt{3}r_gc$ 
(the angular momentum of the marginally stable circular orbit)
then the disk differs drastically from its standard counterpart.
As we discuss below, such a small-scale disk naturally generates hard X-rays
with a break at $\sim 100$~keV. 

The small-scale disk is likely to form in wind-fed sources. Can it also be 
widespread among other observed sources, for instance AGN? In fact, 
the dominant majority of objects with quasi-spherical accretion flows may 
have $l<l_*$ and accrete spherically all the way into the black hole.
It should be difficult to observe such objects because of their
low luminosity (the efficiency of spherical accretion is probably low).
When $l$ exceeds $l_*$ and the disk forms, the luminosity rises dramatically
and the black hole ``switches on'' as an X-ray source. One therefore expects 
to see preferentially the objects with $l$ above the threshold for disk 
formation. If the $l$-distribution of objects falls steeply towards high $l$, 
most of the bright sources should be near the threshold, i.e. the regime 
$\bar{l}_z\sim l_{\rm ms}$ can be widespread among observed {\it bright} 
sources.

%%%%%%%%%%%%%%%%%%%%%%%%%%%%%%%%%%%%%%%%%%%%
\begin{figure}[t]
\begin{minipage}{9.5cm}
\includegraphics[width=90mm]{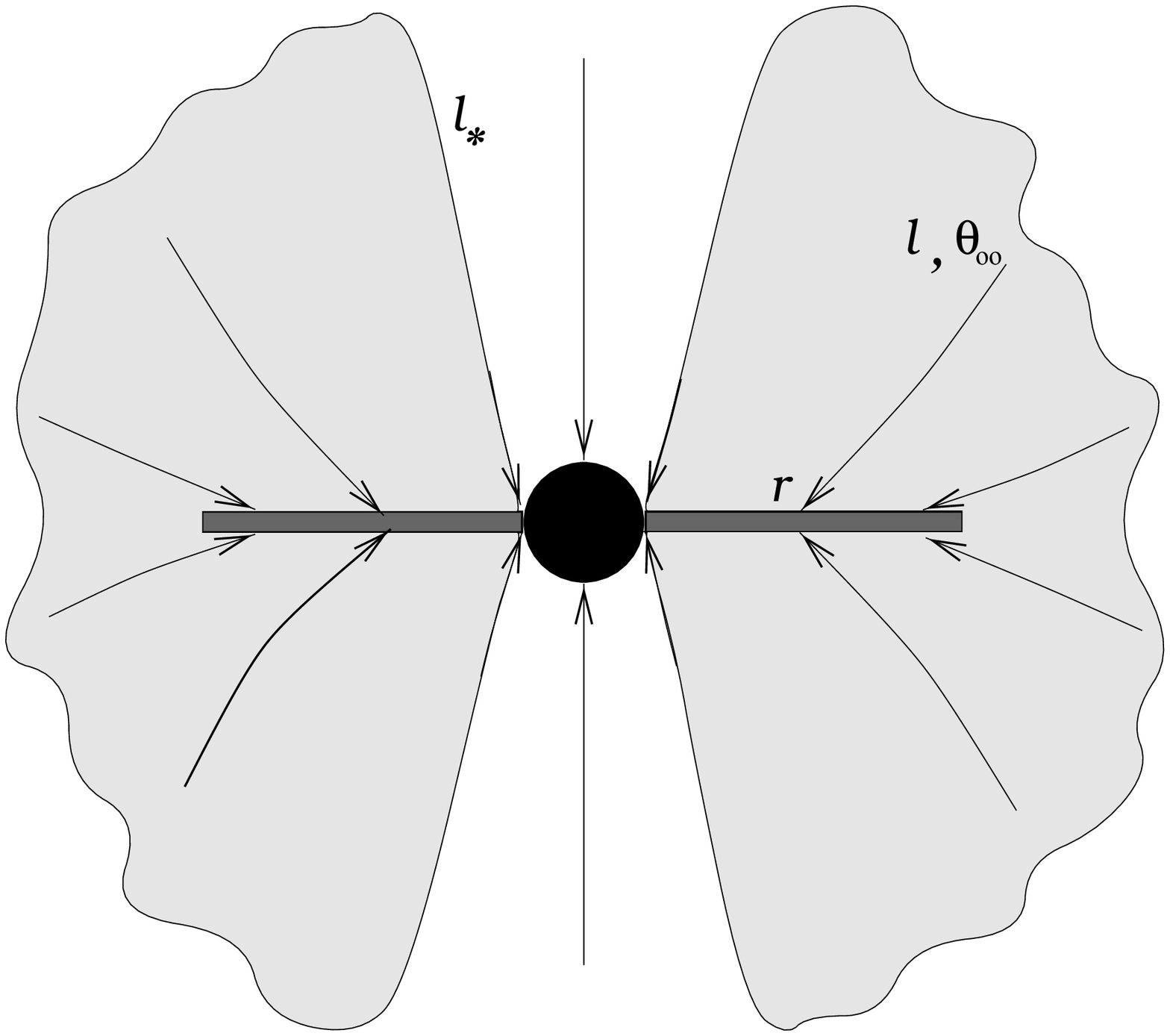}
\end{minipage}
\begin{minipage}{8.5cm}
{\sf Fig.~6. Schematic picture of disk formation.
The shadowed parts of the accretion flow collide
outside $r_g$ and form a couple of radiative shocks which sandwich the
collision ring. The other part of the flow (at small polar angles where
$l<l_*$)
is directly swallowed by the black hole with a low radiative efficiency.}
\end{minipage}
\end{figure}
%%%%%%%%%%%%%%%%%%%%%%%%%%%%%%%%%%%%%%%%%%%%

\subsection*{Sticky Caustic}

For illustration, consider a rotating accretion flow with angular momentum 
$l(\theta_\infty)=l_0\sin\theta_\infty$ (i.e. assume solid body rotation 
at infinity, $\theta_\infty$ is polar angle). 
Free-fall streamlines are symmetric
about the equatorial plane and they intersect in the ring-like caustic
of radius $r_0\approx l_0^2/GM$ (Fig.~6). 
The collisionless shocks enveloping the caustic are pinned to the equatorial
plane if the accretion rate is sufficiently high (BI),
\be
   \dm=\frac{\dM c^2}{\LE}>\dm_0\sim 0.3.
\ee
For the typical efficiency of the small-scale disk (a few percent) this 
condition reads $L\simgt 10^{-2}\LE$. Then the shocked gas forms a thin disk
confined by the infall ram pressure. (By contrast, the regime $\dm<\dm_0$ has 
quasi-spherical shocks that require 2D hydrodynamic simulations, see 
Igumenshchev, Illarionov, \& Abramowicz 1999.) In the regime of pinned shocks,
matter falls freely until it reaches the thin shocked disk.

The vertical structure of the disk consists of a cold thin disk and 
a hot two-temperature layer atop it, where $T_p\gg T_e$.
The infalling matter streams from the shock front downwards, passes through 
the hot layer, cools, and condenses into the cold disk. The electron
temperature  in the hot layer is found from the balance between Compton 
cooling and heating by $e-p$ collisions (BI), 
\begin{equation}
  \frac{kT_e}{m_ec^2}\approx
   \left(\frac{\ln\Lambda\, m_e}{\sqrt{2\pi}m_p}\right)^{2/5}\approx 0.1,
\end{equation}
where $\ln\Lambda\sim 15$ is Coulomb logarithm. 
Thomson optical depth $\tT$ of the layer 
is determined by the time of plasma cooling which yields $\tT\approx 1$.
The cold disk is geometrically thin and turbulent, with very fast vertical 
mixing (BI).  The disk absorbs the infall and shares its momentum. 
In this sense, the disk is a ``sticky'' caustic in the accretion 
flow and it resembles the accretion line of Bondi \& Hoyle (1944).

\subsection*{Inviscid Disk and Its Luminosity}

Accretion flows with $\bar{l}_z<2r_gc$ form disks which can overcome the 
centrifugal barrier and spiral fast into the black hole without any help of 
horizontal viscous stresses. Such an ``inviscid'' disk, however, interacts 
inelastically with the feeding infall and its streamlines are computed from
the laws of momentum and mass conservation (BI), see Fig.~7.
The radius of the inviscid disk can be as large as
$r_{\rm max}=13.6 r_g\approx 27 GM/c^2$; $r_0=r_{\rm max}$ is reached at 
$l_0=\lcr\approx 2.62r_gc$. At $l_0>\lcr$ the centrifugal barrier stops
accretion and the steady inviscid regime is not possible. At $l_0\gg \lcr$
the caustic transforms into the standard viscous accretion disk.

%%%%%%%%%%%%%%%%%%%%%%%%%%%%%%%%%%%%%%%%%%%%
\begin{figure}
\includegraphics[width=85mm,height=85mm]{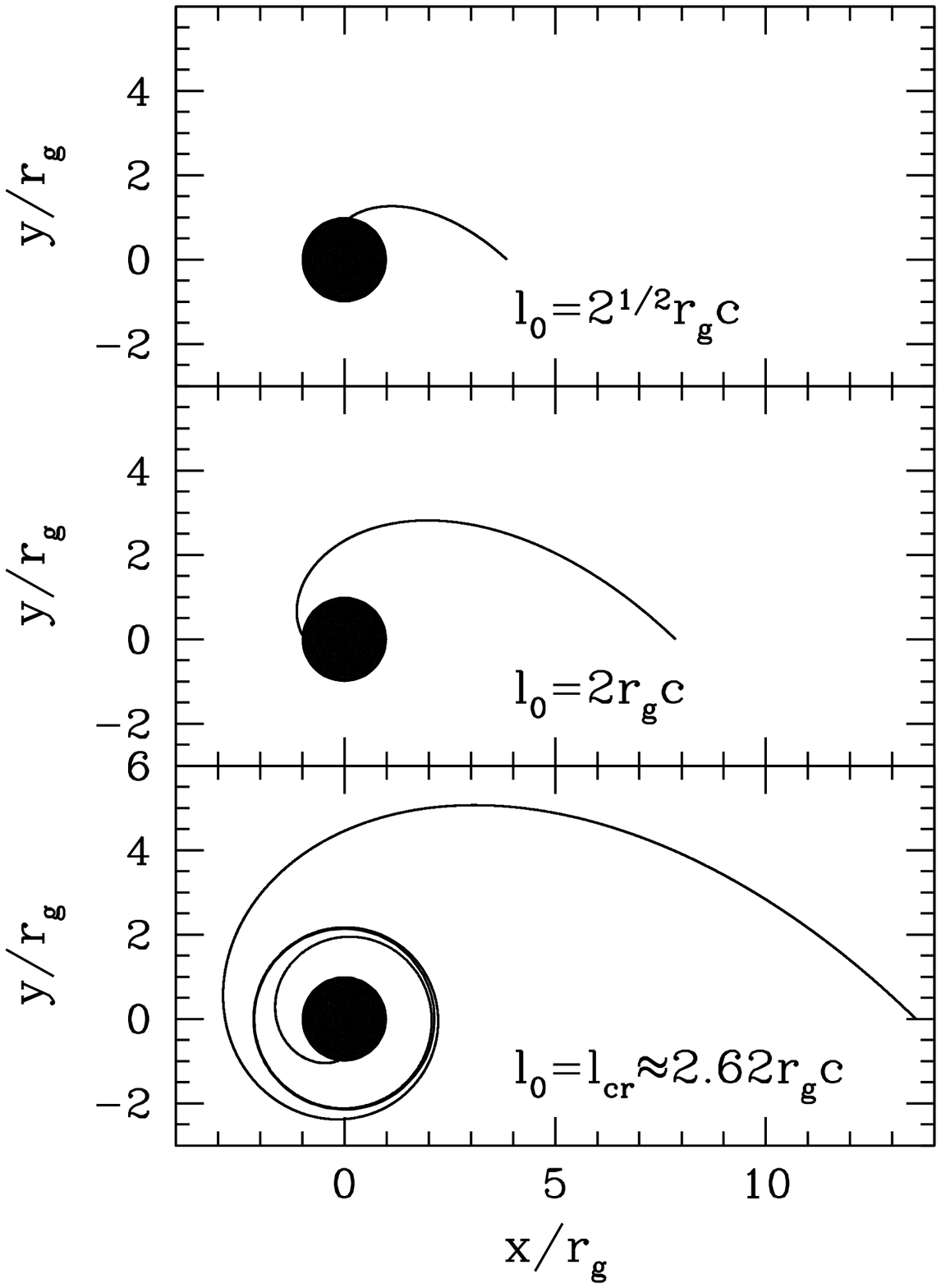}
\includegraphics[width=85mm,height=92mm]{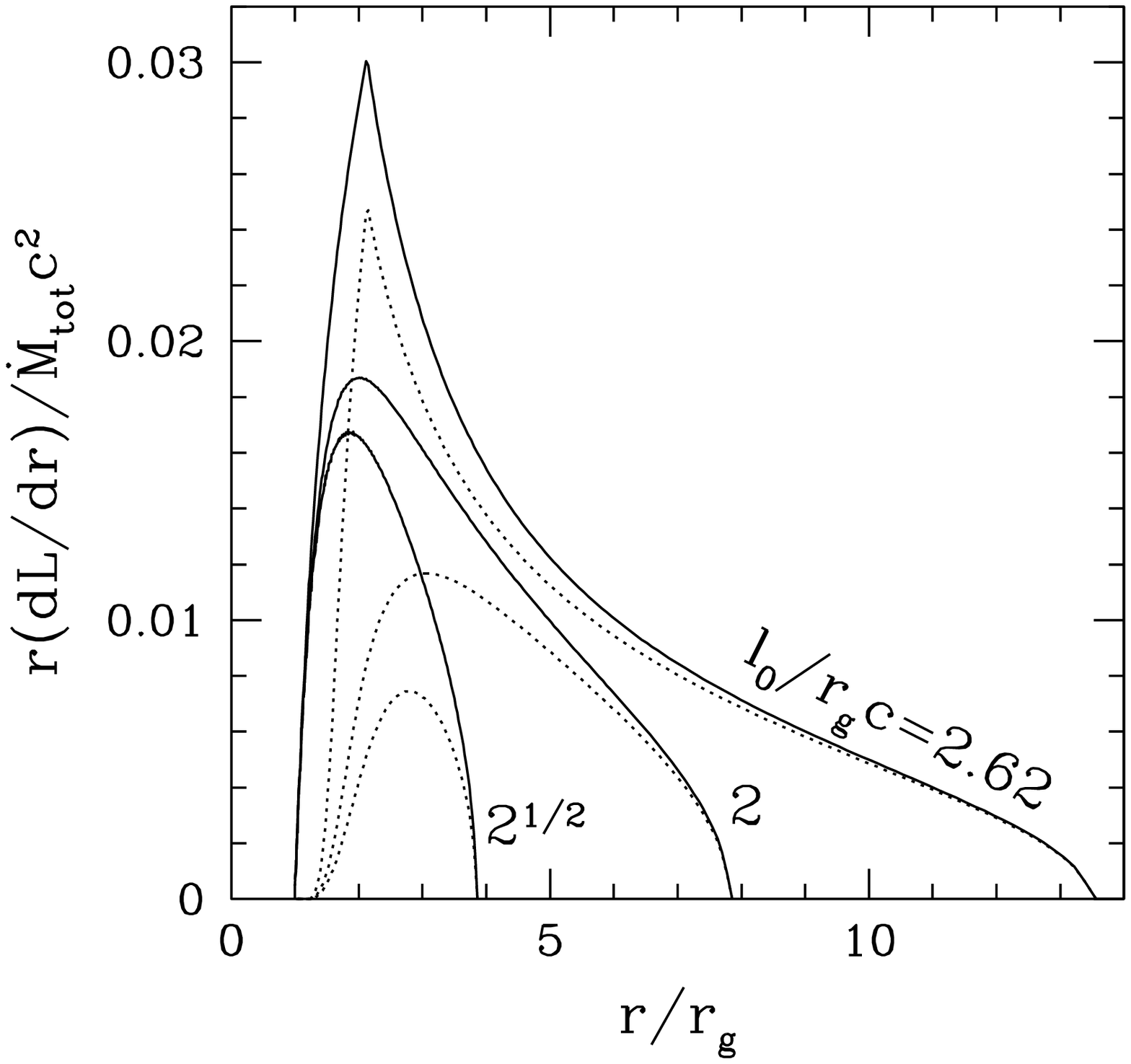}
\begin{minipage}{180mm}
{\sf Fig.~7. {\em (Left)\/} Streamlines of the disk, from the outer edge
into the black hole. The infall has angular momentum 
$l(\theta_\infty)=l_0\sin\theta_\infty$.
In the critical case (bottom panel) the gas makes
infinite number of revolutions at the critical radius $\rcr\approx 2r_g$ 
before it falls into the black hole. {\em (Right)\/}
Radial distribution of the disk luminosity: solid curves --
the released luminosity, dotted curves -- the observed luminosity
corrected for the capture by the black hole.}
\end{minipage}
\end{figure}
%%%%%%%%%%%%%%%%%%%%%%%%%%%%%%%%%%%%%%%%%%%%

The energy released in the infall-disk collision is radiated.
The radial distribution of the luminosity generated by the small-scale 
inviscid disk is shown in Fig.~7. It peaks at $2r_g$ and a substantial 
fraction of the luminosity is captured into the black hole. This capture 
reduces the observed luminosity (the dotted curves in Fig.~7).

\subsection*{X-Ray Spectrum and Variability}

The released energy is emitted partly 
by the hot postshock layer and partly by the cold disk. 
This structure resembles the disk-corona model.
The hot layer is cooled by unsaturated Comptonization producing a power-law 
spectrum with a break at $\sim 100$~keV (see eq.~6), in agreement with 
observations. 
$\Gamma$ and $R$ of the spectrum depend on details to be investigated.
Note that the postshock layer is likely turbulent and inhomogeneous, 
and it may resemble a patchy rather than slab corona.

Fluctuations in the angular momentum 
can cause variations in the size of the small-scale disk 
($r\propto l^2$) and even switch the system to a soft spectral 
state with the standard viscous regime of accretion. 
This may easily occur in wind-fed X-ray binaries since 
the trapped $l$ is sensitive to the wind velocity, $l\propto w^4$, and
may change substantially. The time-scale for $w-$variations
is not known and can be quite long. The flow is also expected to fluctuate 
on $t_*\sim 10$~s, the time of accretion from Compton radius (see 
Illarionov \& Beloborodov 2001).
The $t_*$ may be associated with the peak in the Fourier spectrum of Cyg~X-1. 

Two-component accretion flows (large-scale disk + 
quasi-spherical inflow) should form in X-ray binaries if the massive 
companion fills its Roche lobe; similar flows are also possible in AGN. 
The quasi-spherical component with a maximum angular momentum $l_0$ will 
hit the disk inside the radius $r_0=l_0^2/GM$ and generate 
power-law X-rays; this interaction may also cause transition to 
fast inviscid accretion (Beloborodov \& Illarionov, in preparation). 
Outside $r_0$, the disk may emit according to the standard model.

%############################################################################

\section*{SUPER-EDDINGTON DISKS}

The pattern of accretion at super-Eddington $\dM$ remains uncertain. 
It is sensitive to the unknown vertical distribution of heating $q(z)$. 
Slight increase of $q$ towards upper layers will drive an outflow while
homogeneous (or concentrated to the midplane) heating may allow a stable 
advective inflow. The latter case was studied numerically in the
``slim'' (vertically integrated) approximation in pseudo-Newtonian
(Abramowicz et al. 1988; Chen \& Taam 1995) and Kerr (Beloborodov 1998a) 
gravitational field. 2D simulations 
(Igumenshchev \& Abramowicz 2000) show that different types of flows are 
possible depending on the viscosity/heating prescription.
Future 3D MHD simulations might help to understand the preferred flow pattern
at high $\dM$.

The spectrum of a blackbody pseudo-Newtonian disk was evaluated by 
Szuszkiewicz, Malkan, \& Abramowicz (1996). The blackbody assumption fails
at $\alpha>0.03$: then the thermalization time is longer than the inflow time
and the disk overheats up to $\sim 10^9$~K (Beloborodov 1998a).

%############################################################################

\section*{SUMMARY}

The theory of bright black-hole sources is developing and a number of 
alternative models may account for the data. Three basic possibilities 
suggested for sub-Eddington accretion are: (1) viscous two-temperature disk, 
(2) viscous disk with active MHD corona, and (3) small-scale inviscid disk
in the inner region of a quasi-spherical inflow. 
Different regimes of accretion may take place in different objects, depending 
on the angular momentum of the accreting gas and the accretion rate.
All the three models
produce Comptonized X-ray spectra with a break at $\sim 100$~keV.
In general, an observed Comptonized spectrum is degenerate,
i.e. it is insensitive to the heating mechanism and/or the accretion dynamics.
Detailed studies of the spectrum features such as
the Fe K$\alpha$ line and the Compton reflection bump 
might help to break the degeneracy. Also further studies of source 
variability can provide useful constraints on the models.

%###########################################################################

\end{document}